\documentclass[12pt]{emulateapj}

       \usepackage{graphicx}
       \usepackage{natbib}
\usepackage{graphicx,rotating}
\usepackage{amssymb}
\usepackage{epstopdf}
\usepackage{amsmath}
\usepackage{mathtools}
\usepackage{natbib}
\usepackage{xcolor}
\usepackage{xspace}
\newcommand{\beq}{\begin{equation}}
\newcommand{\eeq}{\end{equation}}
\def\etal{{\sl et~al.~}}
\newcommand{\mAs}{{$\mu$\,Arae~}}
\newcommand{\mA}{$\mu$\,Arae}

\newcommand{\HST}{{\it HST}}
\newcommand{\HSTs}{{\it HST~}}
\newcommand{\FGS}{{\it FGS}}
\newcommand{\FGSs}{{\it FGS~}}
\newcommand{\HIP}{{\it Hipparcos}}
\newcommand{\G}{{\it Gaia}}
\newcommand{\Gs}{{\it Gaia~}}
\newcommand{\kms}{km s$^{-1}$~}
\newcommand{\kmse}{km s$^{-1}$}
\newcommand{\ms}{m s$^{-1}$}

\newcommand{\msini}{$\cal{M}$\,sin\,{i}~}

\newcommand{\m}{$\cal{M}$}

\newcommand{\mjupe}{$\cal{M}_{\rm Jup}$}

\newcommand{\msune}{$\cal{M}_{\odot}$}

\newcommand{\br}{\rm b}

\newcommand{\dr}{\rm d}
\newcommand{\er}{\rm e}

\def\vplanet{\texttt{\footnotesize{VPLanet}}\xspace}
\def\spinbody{\texttt{\footnotesize{SpiNBody}}\xspace}

\def\HillStability{\texttt{\footnotesize{HillStability}}\xspace}

\received{}
\revised{} 
\accepted{}

\shorttitle{The \mAs System}
\shortauthors{Benedict \etal}

\begin{document}
\bibliographystyle{/Active/my2}

\title{
The \mAs Planetary System: Radial Velocities and Astrometry}

\author{G. F. Benedict}
\affiliation{McDonald Observatory, University of Texas, Austin, TX 78712}
 \author{B. E.  McArthur}
\affiliation{McDonald Observatory, University of Texas, Austin, TX 78712}
\author{E. P. Nelan}
\affiliation{Space Telescope Science Institute, Baltimore, MD 21218}
\author{R. Wittenmyer}
\affiliation{Centre for Astrophysics, University of Southern Queensland, Toowoomba Qld 4350, Australia}
\author{R. Barnes}
\affiliation{University of Washington, Seattle, WA 98195}
\author{H. Smotherman}
\affiliation{University of Washington, Seattle, WA 98195}
\author{J. Horner}
\affiliation{Centre for Astrophysics, University of Southern Queensland, Toowoomba Qld 4350, Australia}



\begin{abstract}
With {\it Hubble Space Telescope} Fine Guidance Sensor astrometry and published and previously unpublished radial velocity measures we explore the exoplanetary system \mA . Our modeling of the radial velocities results in improved orbital elements for the four previously known components. Our astrometry contains  no evidence for any known companion, but provides upper limits for three companion masses.   A final summary of all past Fine Guidance Sensor exoplanet astrometry results uncover a bias towards small inclinations (more face-on than edge-on). This bias remains unexplained by either small number statistics,  modeling technique,  Fine Guidance Sensor mechanical issues, or orbit modeling of noise-dominated data. A numerical analysis using our refined orbital elements suggests that planet d renders the \mAs system dynamically unstable on a timescale of $10^5$ years, in broad agreement with previous work.\end{abstract}





%

\section{Introduction}

Multiple planet systems provide an opportunity   to probe the dynamical origins of planets \citep[e.g.][]{For06}. Every multiple planet system has the potential to serve as a case study of planetary system evolution \citep{Wri09}. They provide laboratories within which to tease out the essential processes and end states from the accidental. \mAs is such a system. 

The \mAs system  is one of the best known multi-planet systems, with components having received official IAU names in late 2015. \cite{But01} announced the discovery of \mA\,b, which was initially thought to move on an eccentric orbit. \cite{Pep07}  presented new observations of the \mAs system, revealing the four components known today. Using Doppler spectroscopy, that team announced  the discovery of component c and firmed up the period of component e. This multiplanet system has until now, only minimum masses for the four components (with periods   $9.6< P < 3900$ days, Pepe et al. 2007\nocite{Pep07}). With  access to only radial velocity observations (RV) the inferred masses depend on their orbital inclination angle, $i$, providing  minimum mass values, $0.03 < {\rm \cal{M}} \sin i< 1.8$\mjupe, for the four companions found by RV. Hence, we included this system in a {\it Hubble Space Telescope} (\HST) proposal \citep{Ben07c} to carry out astrometry using the Fine Guidance Sensors (\FGS). Those observations supported attempts to establish true component mass and the architectures of several promising candidate systems, all relatively nearby, and with companion \msini values and periods suggesting measurable astrometric amplitudes.  

For \mAs we follow analysis procedures previously employed for the exoplanetary systems $\upsilon$ And \citep{McA10}, HD 136118 \citep{Mar10}, HD 38529 \citep{Ben10}, HD 128311 \citep{McA14}, and HD 202206 \citep{Ben17b}. 
\mAs companion masses and the \mAs system architecture were our ultimate goals.  Unfortunately, our astrometric investigation of \mA, yields only a parallax  consistent with the \Gs EDR3 values. Based on the astrometric residual statistics, we estimate upper mass limits for components \mA\,b, d, and e. These limits are consistent with both the \Gs precision and the lack of acceleration obtained from a comparison of \HIP~ and \Gs EDR3 proper motions \citep{Bra21}.

Section~\ref{Ap1} identifies the sources of RV and our modeling results. Section~\ref{AST} describes the astrometric data and modeling techniques used in this study. After determining parallax and proper motion, we subject the residuals to periodogram analysis and find no significant signals at any of the periods determined from the RV (Section~\ref{nada}).  Our astrometric precision yields only upper limits on possible companion masses. We discuss these results  in comparison to past FGS astrometric results (Section~\ref{Disc}), and briefly revisit system stability in Section~\ref{Disc1}. Lastly, in Section~\ref{Summ} we summarize our findings. 

Table~\ref{tbl-STAR} contains previously determined information and sources for the host star subject of this paper, \mA. 
We abbreviate millisecond of arc as mas throughout and state times as mJD=JD-2400000.

\section{\mAs Radial Velocities}  \label{Ap1}

\cite{Pep07} reported previous and new  radial  velocities, components of the stellar orbital motion around the barycenter of the system, with Doppler spectroscopy.  We list all RV data with sources in Table~\ref{tbl-newRV}. We take the CORALIE RVs from \cite{Pep07}.  To these we add new 
publicly-available data from the HARPS spectrograph on the 3.6m ESO 
telescope at La Silla \citep{Tri20}.    We also include 180 
RV measurements from the UCLES 
spectrograph \citep{Die90} on the 3.9m Anglo-Australian Telescope, 
gathered as part of the 18-year Anglo-Australian Planet Search program 
\citep[e.g.][]{Tin01, Wit14, Wit17}.  For all data sets, where there 
were multiple observations in a single night, we binned them together 
using the weighted mean value of the velocities in each night. We 
adopted the quadrature sum of the rms about the mean and the mean 
internal uncertainty as the error bar of each binned point.

\begin{figure}
\includegraphics[width=6.5in]{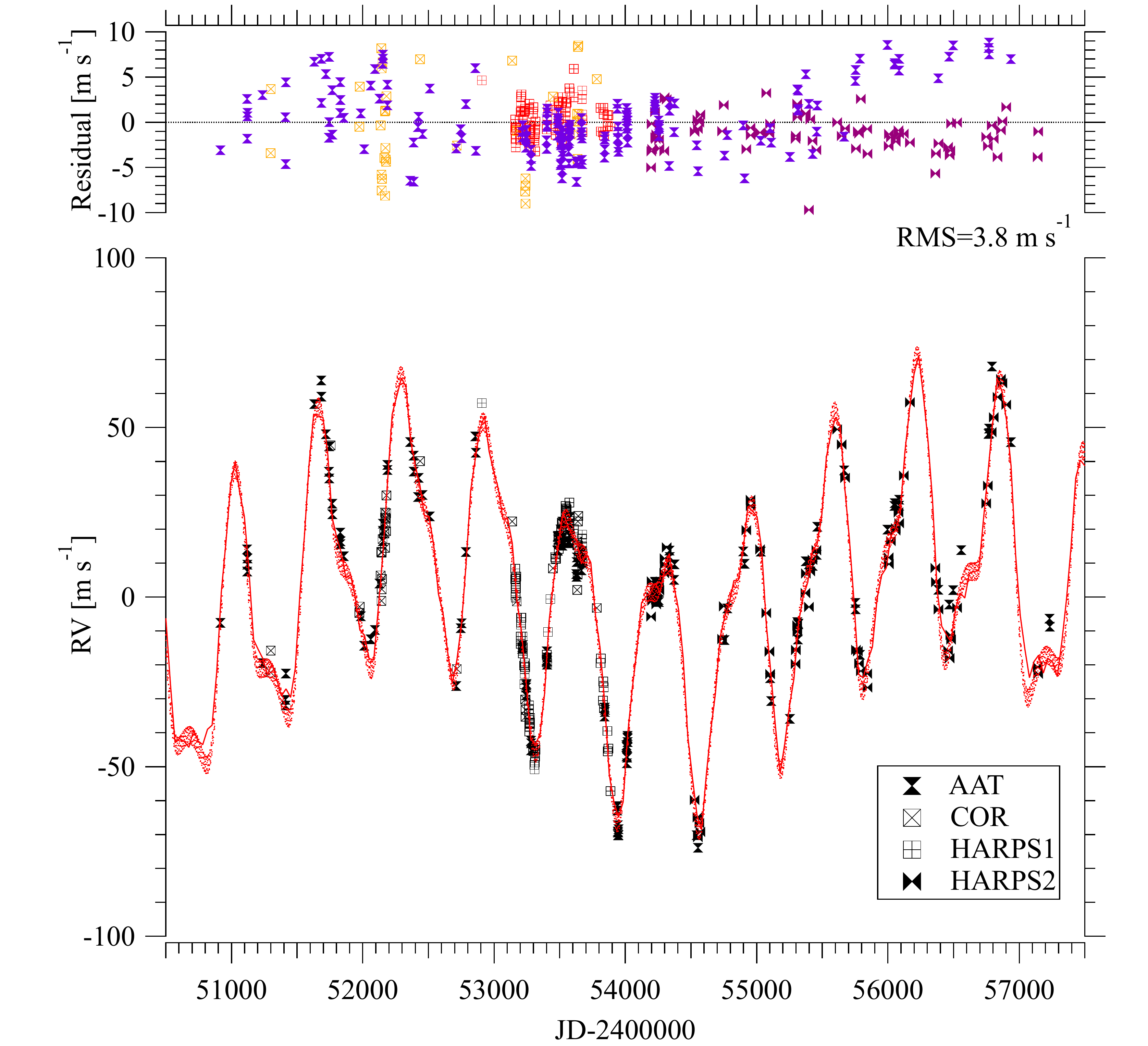}
\caption{RV values from the sources listed in Table~\ref{tbl-newRV} plotted on  the final RV four component orbit (Table~\ref{tbl-RVORB}). All RV input errors have been increased by a factor of 1.4 to achieve a near unity $\chi^2$. Residuals are plotted in the top panel. We note the RMS RV residual value in the plot.} \label{fig-RVf}
\end{figure}

This changing velocity, $v$, is the projection of a Keplerian orbital velocity to the observer's line of sight plus a constant velocity, $\gamma$. $K$ is the velocity semi-amplitude in \kmse. The total RV signal  we model includes contributions from all components. Because our GaussFit modeling results critically depend on the input data errors, we first modeled the RV to assess the validity of the original input RV errors.  In order to achieve a $\chi^2/DOF$ of unity for our solution required increasing the original errors on the RVs  by a  factor of 1.4 for CORALIE and UCLES, and by 2.0 for  HARPS. This suggests that either the errors were underestimated, or that  that the fit is not as good as it could be (i.e. evidence that there may be more to learn about the system). Figure~\ref{fig-RVf} presents RV plotted as a function of time and the final combined orbital solution. The RMS residual is 3.8 \ms. Table~\ref{tbl-gammas} contains derived velocity offsets for each RV source. Table~\ref{tbl-RVORB} contains orbital elements and $1-\sigma$ errors for components b, c, d, and e based on these RV.

 \mAs has always presented stability challenges \citep{Pep07, Tim13, Las17,Agn18}. 
Given the frequency with which intrinsic stellar activity has been found to mimic a Keplerian signal in RV data \citep[e.g.][]{Rob14, Rob15, Raj16, Dia18}, we examined the available activity indicators from the HARPS spectra for $\mu$ Ara to determine whether or not all RV signals are dynamical not stellar activity.  

We obtained the complete set of activity indicators from the recently released HARPS RVBANK \citep{Tri20}, which has corrected for nightly zero-point offsets and other systematics.  The available indicators are: FWHM, bisector, H$\alpha$, and the two Na D lines.  Using the online \textit{Agatha} tool\footnote{\url{https://phillippro.shinyapps.io/Agatha}} \citep{Fen17}, we computed four periodograms (Bayes Factor, Maximum Likelihood, Bayesian Generalised Least Squares, and Generalised Least Squares) for each of these activity-indicator time series to search for activity-related signals.  The only significant periodicities were those near one year (357-368 days), with the bisectors alone showing a significant peak at 497 days.  Thus, none of the  RV signals attributed to $\mu$ Ara companions can be attributed to  line profile distortion due to stellar activity. 


\section{\mAs Astrometry}\label{AST}
Unless otherwise noted, for \mAs we carried out  exactly the same analysis detailed in \cite{Ben17b} for HD\,202206.
\subsection{Astrometric Data}
For this study astrometric measurements came from Fine Guidance Sensor\,1r (\FGS\,1r), an upgraded \FGSs installed in 1997 during the second \HSTs servicing mission. 

\begin{center}
\begin{figure}
\includegraphics[width=6in]{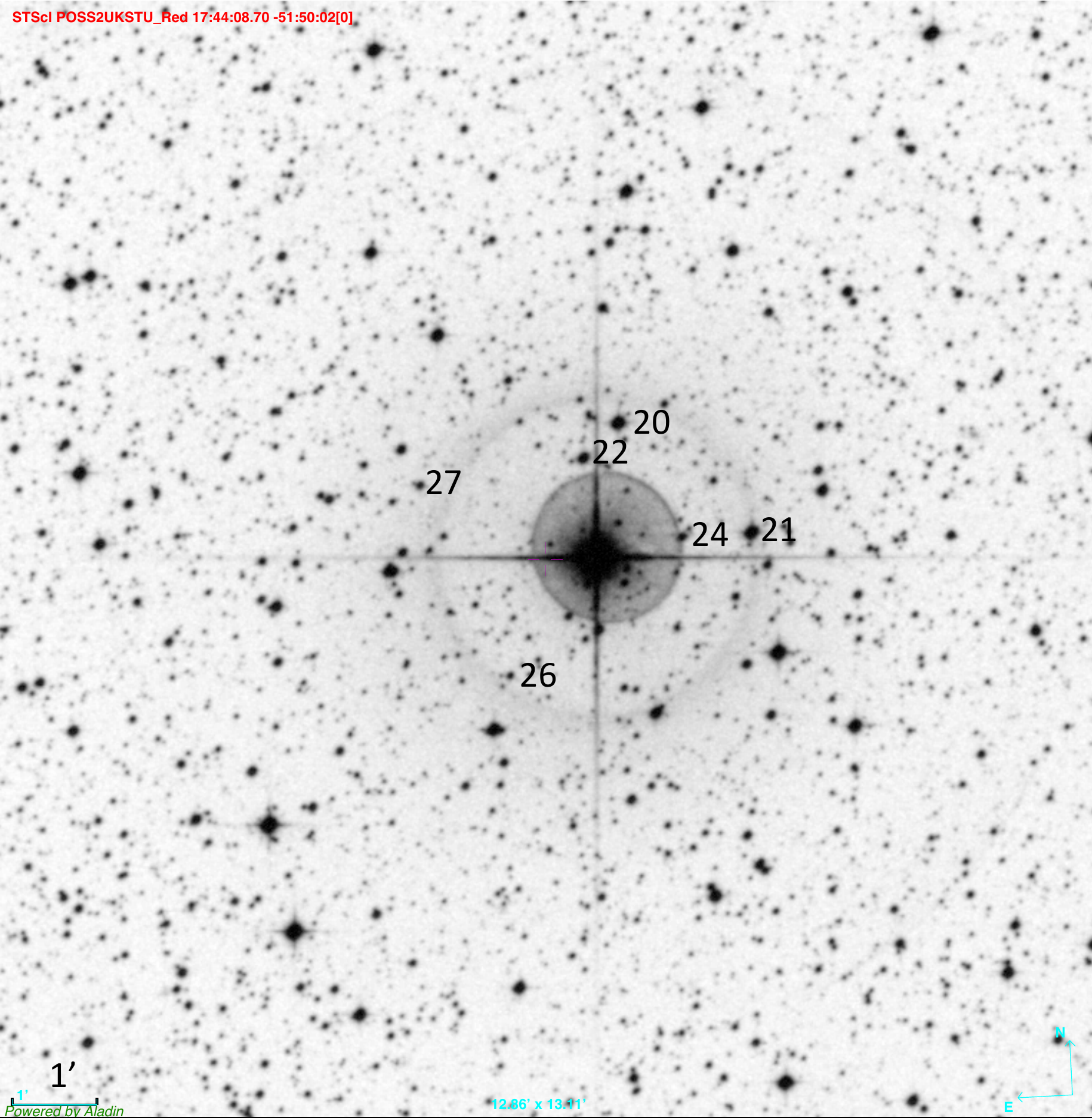}
\caption{\mAs and the astrometric reference stars (20-27) identified in Table~\ref{tbl-piref}. }
\label{fig-Find}
\end{figure}
\end{center}

\begin{center}
\begin{figure}
\includegraphics[width=6in]{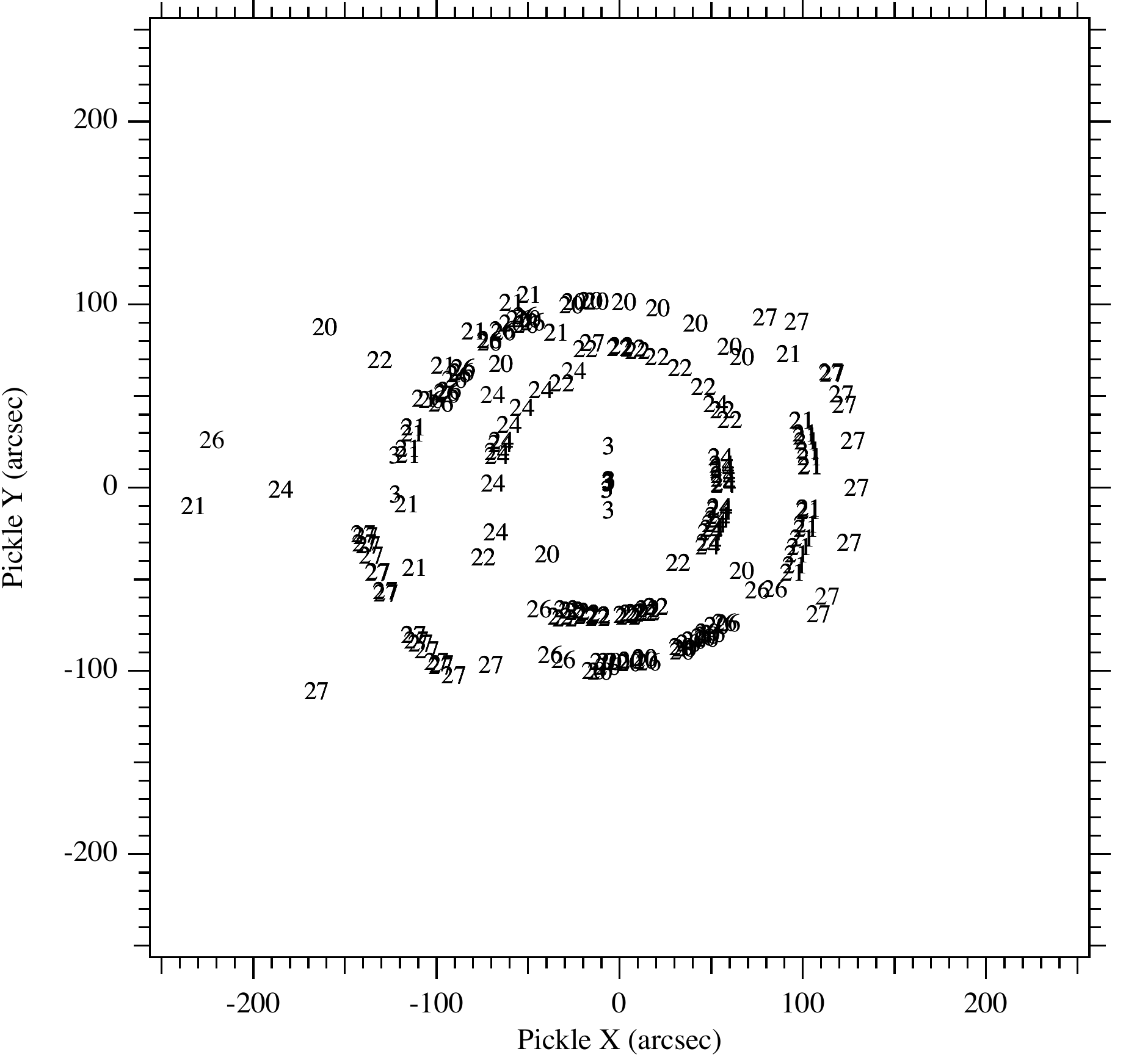}
\caption{Positions of \mAs (3) and astrometric reference stars (20 -- 27) in \FGS\,1r \textbf{FOV} coordinates.}
\label{fig-Pick}
\end{figure}
\end{center}

We utilized only the fringe tracking mode (POS-mode; see Benedict et al. 2017 for a review of this technique, and Nelan et al. 2015 for further details)\nocite{Nel15a,Ben17} in this investigation. POS mode observations of a star 
have a typical duration of 60 seconds, during which over two thousand individual position measurements are collected. We estimate the astrometric centroid  by choosing the median measure, after filtering large outliers (caused by cosmic ray hits and particles trapped by the Earth's magnetic field). The standard deviation of the measures provides a measurement error. We refer to the aggregate of astrometric centroids of each star secured during one visibility period as an ``orbit". We identify the astrometric reference stars and science target in Figure~\ref{fig-Find}. Figure~\ref{fig-Pick} shows the final measured location pattern within \FGS\,1r. 

We present a complete ensemble of time-tagged  \mAs and reference star astrometric measurements, OFAD\footnote{The Optical Field Angle Distortion (OFAD) calibration {\citep{McA06}}.}- and intra-orbit drift-corrected, in Table~\ref{tbl-DATA}, along with calculated parallax factors in Right Ascension and Declination. These data, collected from 2007.5 to 2010.4,  in addition to providing material for confirmation of our results, could ultimately be combined with \Gs measures to significantly extend the time baseline of the astrometry, thereby improving proper motion and perturbation characterization.
\subsection{Astrometry Modeling Priors}\label{CORR}
As in all 
of our previous \FGSs astrometry projects, e.g., \cite{Ben01,Ben07, Ben11, Ben16, Ben17b} and \cite{McA10, McA11} we include as much prior information as possible in
our modeling.    We utilize parallax, proper motion, cross-filter, and lateral color 
calibration priors in this analysis. 

Past investigations 
\citep[e.g.][]{Har99,Ben11} derived reference star parallaxes from a combination of photometry and spectroscopy. In support of this approach we obtained spectroscopy of the reference stars,  long before the publication of \Gs EDR3. We used the RC Spectrograph on the CTIO Blanco 4m. 
The Loral 3K CCD detector with KPGL1-1 grating delivered a dispersion of 1.0 \AA ~pixel$^{-1}$, covering the wavelength
range 3500\AA$<\lambda < 5830$ \AA.
Classifications used a combination of template matching and line ratios. We estimate spectral types (included in Table~\ref{tbl-piref} for completeness)  with precision generally better than $\pm2$ subclasses. 

\begin{center}
\begin{figure}
\includegraphics[width=5in]{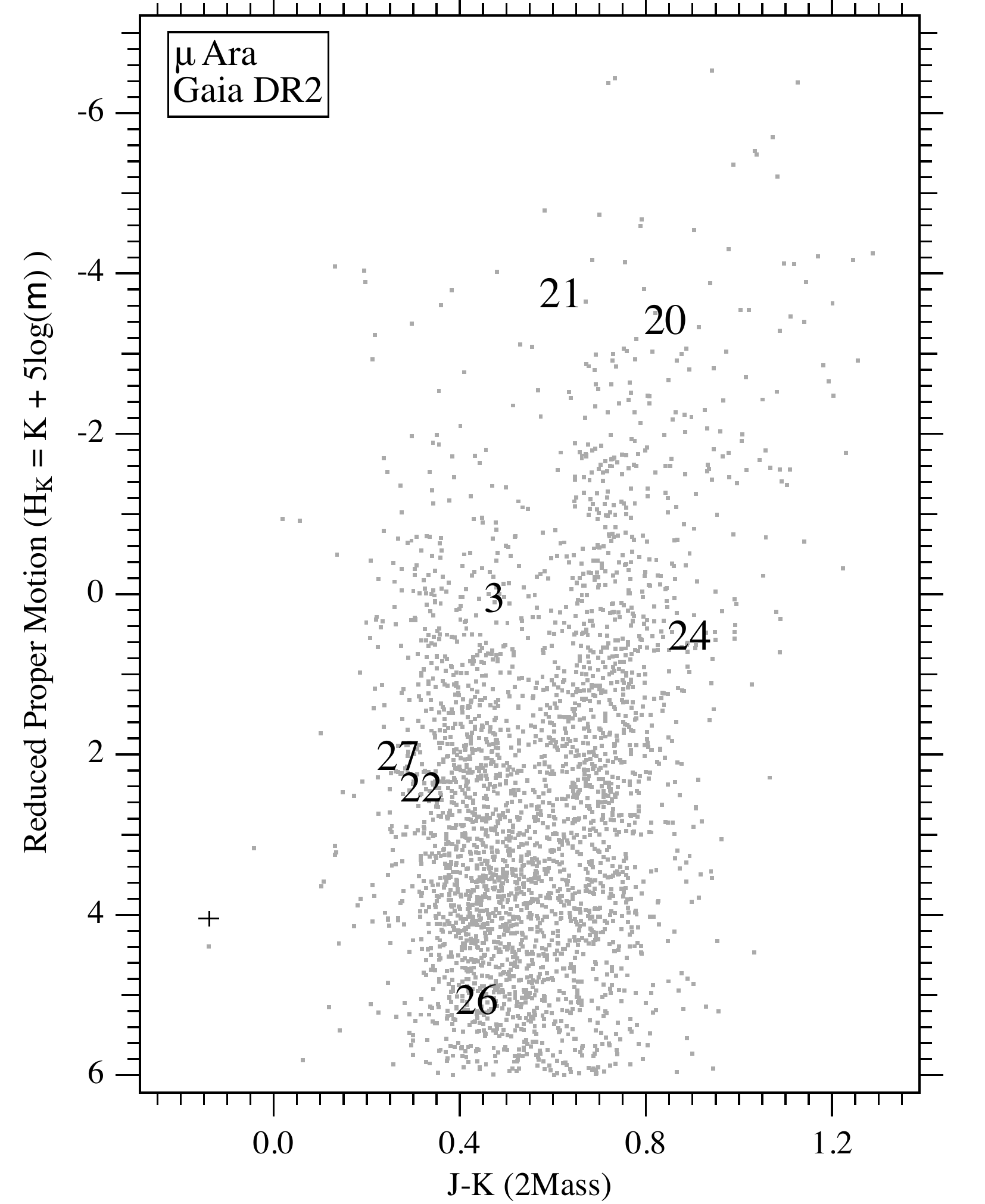}
\caption{Reduced proper motion diagram for 3200 stars in a 1\arcdeg ~field centered
on \mA, \#3 on the plot.  Star identifications are in Table~\ref{tbl-piref} and proper motions are from \Gs DR2. 
For a given spectral type, 
giants and sub-giants have more negative $H_K$ values and are redder than dwarfs in 
$(J-K)$.  The small cross at the lower left represents a typical $(J-K)$ error of 0.04 mag and $H_K$ error of 0.17 mag.  
} 
\label{fig-RPM}
\end{figure}
\end{center}

To check the luminosity classes obtained from classification spectra and the \Gs EDR3 parallaxes \citep{Gia21} we obtain  proper 
motions from the 
EDR3 for a one-degree-square field centered on \mA, and then 
produce a reduced proper motion diagram \citep{Str39,Yon03,Gou03}  as additional confirmation. 
 Figure~\ref{fig-RPM} contains the reduced proper motion diagram for the \mAs field, including \mAs and our reference stars. 
We employ the following priors;
\begin{enumerate}

\item \textbf{Parallax:} 
Rather than rely on spectrophotometric reference star parallax estimates, 
this investigation simply adopts EDR3 
values \citep{Gia21}. It should be noted, however, that we do not treat those values as being hardwired or absolute. Instead, we consider them to be quantities (Table~\ref{tbl-piref}) introduced as observations with error. The average EDR3 parallax error is 0.02 mas. We also list the $RUWE$ (Renormalised Unit Weight Error) for each reference star. \cite{Sta21} find that the \Gs $RUWE$  robustly predicts unmodeled photocenter motion, even in the nominal "good" range of 1.0--1.4 \citep[see also][]{Bel20}. To test the effect of such tight priors, the results presented below include two separate runs of the model: the first with the original EDR3 errors, the second with uniform 1.0 mas parallax errors on those priors. Note that we utilize no parallax prior for \mA, an independent parallax having some value.
\item \textbf{Proper Motions:} For the reference stars we use proper motion priors from EDR3. 
Simply relying on the EDR3 values for the reference stars might introduce a bias, given the limited EDR3 time span and the potentially complicated perturbations from the known components.  Again,  we present the two model run results below, the first including the original EDR3 proper motion priors errors, averaging 0.02 mas yr$^{-1}$, the second increasing the proper motion priors errors to 1.0 mas yr$^{-1}$. Again, we utilize no proper motion priors for \mA.

\item \textbf{Lateral Color Corrections:}  These corrections, entered into the model as data with errors, are identical to those used in \cite{Ben17b}.

\item \textbf{Cross-Filter Corrections:}
{\it \FGS\,1r} contains a neutral density filter, reducing the brightness of \mAs by five magnitudes (from $V=5$ to $V=10$), permitting us to relate the measured positions of \mAs  to far fainter reference stars 
all with $V>12$. While every effort is made to build filters with plane-parallel surfaces, they are not, so some positional shift is introduced between filter-in and filter-out measures. Section 2 of \cite{Ben02a} describes how we derive this correction for \FGS\,3. Our measured values for \FGS\,1r 
were  $\Delta XF_x=8.15\pm0.14$ mas and $\Delta XF_y=-0.66\pm0.21$ mas, again, quantities introduced as observations with error in the model shown below. 
\end{enumerate}

\subsection{Modeling the \mAs Astrometric Reference Frame}  \label{AstRefs}
The astrometric reference frame for \mAs consists of six stars (Table~\ref{tbl-piref}). 
The \mAs field (Figure~\ref{fig-Find}) exhibits the distribution of  astrometric reference stars (ref-20 through ref-27) used in this study. 
The \mAs field was observed at a very limited range of spacecraft roll 
values (Table~\ref{tbl-DATA}). 
Figure \ref{fig-Pick} shows the distribution in \FGS\,1r  coordinates of the thirty-two
 sets (epochs) of \mAs and  reference star measurements.    We placed \mAs (labeled 3)  in several different y locations within the \FGS\,1r total field of view (FOV)
to maximize the number of astrometric reference stars  and to ensure guide star availability for the other two \FGSs units. 
At each epoch we measured each reference star 1--4 
times, and \mAs 3--5 times. 

 Our choice of model (Equations 3-4) was driven entirely by the goodness of fit for the reference stars. We used no \mAs observations 
to determine the reference frame mapping coefficients, $A--F$. Depending on astrometer (\FGS\,1r) and telescope (\HST) distortions, we can solve  for 
\begin{enumerate}
\item roll, offsets,  and global scale (4c, where we substitute $-B$ for $D$ and $A$ for $E$ in Equation 4, with $R_x$, $R_y$ removed from Equations 3-4), 
\item roll, offsets, independent scales along each axis (6c, Equations 3-4)
\end{enumerate}
By changing from 4c to 6c, 
we suffer a 6\% loss in degrees of freedom (DOF) but obtain a 35\% reduction in $\chi^2$/DOF.

\subsection{The  Model} \label{myMod}
From  
positional measurements we determine the scale, rotation, and offset ``plate
constants" relative to an arbitrarily adopted constraint epoch for
each observation set. We employ GaussFit (Jefferys \etal 
1988) 
\nocite{Jef88} to minimize $\chi^2$. The solved equations of condition for the 
\mAs 
field are:
\begin{equation}
        x^\prime = x + lc_x(\it B-V) - \Delta XF_x
\end{equation}
\begin{equation}
        y^\prime = y + lc_y(\it B-V)  - \Delta XF_y
\end{equation}

\begin{equation}
\xi  = Ax^\prime + By' + C  \\- \mu_\alpha \Delta t  - P_\alpha\varpi - \sum_{n=1}^{4} O_{n,x} 
\end{equation}
\begin{equation}
\eta = Dx^\prime + Ey^\prime + F  \\- \mu_\delta \Delta t  - P_\delta\varpi - \sum_{n=1}^{4} O_{n,y}
\end{equation}

\noindent 
Identifying terms, $\it x$ and $\it y$ are the measured coordinates from {\it HST};   $(B-V)$ is the Johnson $(B-V)$ color of each star; and $\it lc_x$ and $\it lc_y$ are the lateral color corrections, $\Delta XF_x$ and $\Delta XF_y$ are cross filter corrections applied only to \mA.  $A$, $B$, 
$D$, $E$,  
are scale and rotation plate constants, $C$ and $F$ are offsets; $\mu_\alpha$ and $\mu_\delta$ are proper motions; $\Delta t$ is the time difference from the constraint plate epoch; $P_\alpha$ and $P_\delta$ are parallax factors;  and $\it \varpi$ is  the parallax.   $O_x$ and $O_y$ are functions of the classic orbit parameters: $\alpha$, the perturbation semi major axis, $i$, inclination, $e$, eccentricity, $\omega$, argument of periastron, $\Omega$, longitude of ascending node, $P$, orbital period, and $T_0$, time of periastron passage for each included component  \citep{Hei78,Mar10}. $\xi$ and $\eta$ are 
relative positions in RA and Dec that (once scale, rotation, parallax, the proper motions 
and the $O$ are determined) should not change with time.

We obtain the parallax factors from a JPL Earth orbit predictor 
(Standish 1990)\nocite{Sta90}, version DE405. We obtain an orientation to the sky for the 
\FGS\,1r constraint plate (set 18 in Table~\ref{tbl-DATA}) from ground-based astrometry (the UCAC4 Catalog) with uncertainties 
of $0\fdg06$. At this stage we model $only$ astrometry and $only$ the reference stars. 
 From 
histograms of the  reference frame model 
astrometric residuals (Figure~\ref{fig-FGSH}) we conclude 
that we have a well-behaved reference frame solution exhibiting residuals with Gaussian distributions with dispersions $\sigma_{(\rm x,y)} = 1.2,1.1$ mas. The  reference frame 
'catalog' from  \FGS\,1r in $\xi$ and $\eta$ standard coordinates (Table \ref{tbl-POS}) 
was determined with average uncertainties, $\langle\sigma_\xi \rangle= 0.70$ and $\langle\sigma_\eta \rangle = 0.57$ mas. Because we have rotated our constraint plate to an RA, DEC coordinate system, $\xi$ and $\eta$ are RA and DEC.


 At this step in the  analysis the astrometry knows nothing of the RV detections (Table~\ref{tbl-RVORB}). With our derived  $A$, $B$, 
 $D$, $E$, 
 $C$, and $F$ 
 we transform the \mAs astrometric measurements, applying $A$ through $F$ as constants, solving only for \mAs proper motion and parallax, using no priors for \mA.  Table~\ref{tbl-SUM}  compares values for the parallax and proper motion of \mA 
~from {\it HST} and \G~\citep{Gia21}. While the parallax values agree within their respective errors, we note a  disagreement in the  proper motion vector ($\vec{\mu}$) absolute magnitude and direction. This could be explained both by our non-global proper motion measured against a small sample of reference stars, and the limited duration of both astrometric studies, possibly affected by the companion perturbations. 
Alternatively, the mismatch between our proper motion, established through measurements taken from 2007.5 to 2010.4, and the \Gs EDR3 value, a result of a campaign spanning 2014.6 - 2017.4, could indicate acceleration due to the companions. These differences are $\Delta \mu_{\rm RA}$= +0.65 mas yr$^{-1}$ and $\Delta \mu_{\rm Dec}$= +0.62 mas yr$^{-1}$. Table~\ref{tbl-mus} lists the proper motion difference between our model results with weaker proper motion priors and \Gs EDR3, showing two reference stars, ref-20 and ref-22, with differences almost as large as those for \mAs. Furthermore, \cite{Bra21} finds a low $\chi^2$ value when solving a model assuming no proper motion change, comparing \HIP~ with \G, indicating little to no \mAs acceleration over a roughly 25 year time span.

 \section{Astrometric Detection Limits for \mAs Companions} \label{nada}
 We included \mAs in our original \HSTs proposal based on an expected perturbation ($2\times\alpha$) for each minimum mass (\msini) companion, obtained through {\rm pert} = $0.2 (P^{2/3}$\m$_p$)/((d/10)$\times$ \m$_*^{2/3}$) {\rm mas},
with $P$ the companion period, \m$_p$ the known \msini, d the distance in pc, and \m$_*$ the mass of \mA. The then known minimum masses (little changed by our Table~\ref{tbl-RVORB} improved orbits) were  
\msini  b,c,d,e = 1.7, 0.03, 0.52, 1.81 \mjupe, yielding minimum perturbation sizes 0.28, 0.0003, 0.05, 0.99 mas. Clearly,  \FGS~ astrometry had no hope of detecting \mA\,c, but the \HSTs Time Allocation Committee agreed it was worth a shot for at least two of the other components, b and e.
As previously mentioned, the reference frame solution exhibited residual Gaussian distributions with dispersions $\sigma_{(\rm x,y)} = 1.2,1.1$ mas. The \mAs residuals have $\sigma_{(\rm x,y)} = 1.7,1.5$ mas, possibly signalling unmodeled motion.
These residuals should now contain only measurement noise, possible systematic effects,  and perturbations due to suspected companions, \mAs b, c, d, and e. 
 
 We now have access to other predictive resources. These include the \Gs EDR3 $RUWE$ parameter which  predicts unmodeled photocenter motion \citep{Sta21}, and the \cite{Bra21} $\chi^2$ value. The latter parameter measures an amount of measured acceleration obtained by comparing an earlier epoch proper motion from \HIP~with an EDR3 proper motion. A larger $\chi^2$ value indicates more significant  change (acceleration) in proper motion, thus a higher probability of a perturbing companion. Table~\ref{tbl-incs} lists results from all past
 \FGS~exoplanet astrometry, carried out to establish companion masses.  For each result we  tabulate $RUWE$ and degree of likely acceleration, given by the  $\chi^2$ value.  The entries are sorted by $RUWE$ value,  highest to lowest, more potential unmodeled (by \G) image motion to less. Note that the subject of this study, \mA, sits at the bottom. Neither $RUWE$ nor the relatively low $\chi^2$ value predict ease of companion detection. Higher values might be caused by the still experimental \Gs centroiding for bright stars. To test this possibility, we  sampled  24 stars, $3.6<G<7$, within $4\arcdeg$ of \mA. This sample had median $RUWE$, $\chi^2$ values of 1.0 and 5.4, giving \mA, with 0.86 and 2.35,  a low probability of companion detectability. Note that $\gamma$ Cep AB is a long-period binary star system, hence the very large  $\chi^2$ value. 

\begin{figure}
\includegraphics[width=6in]{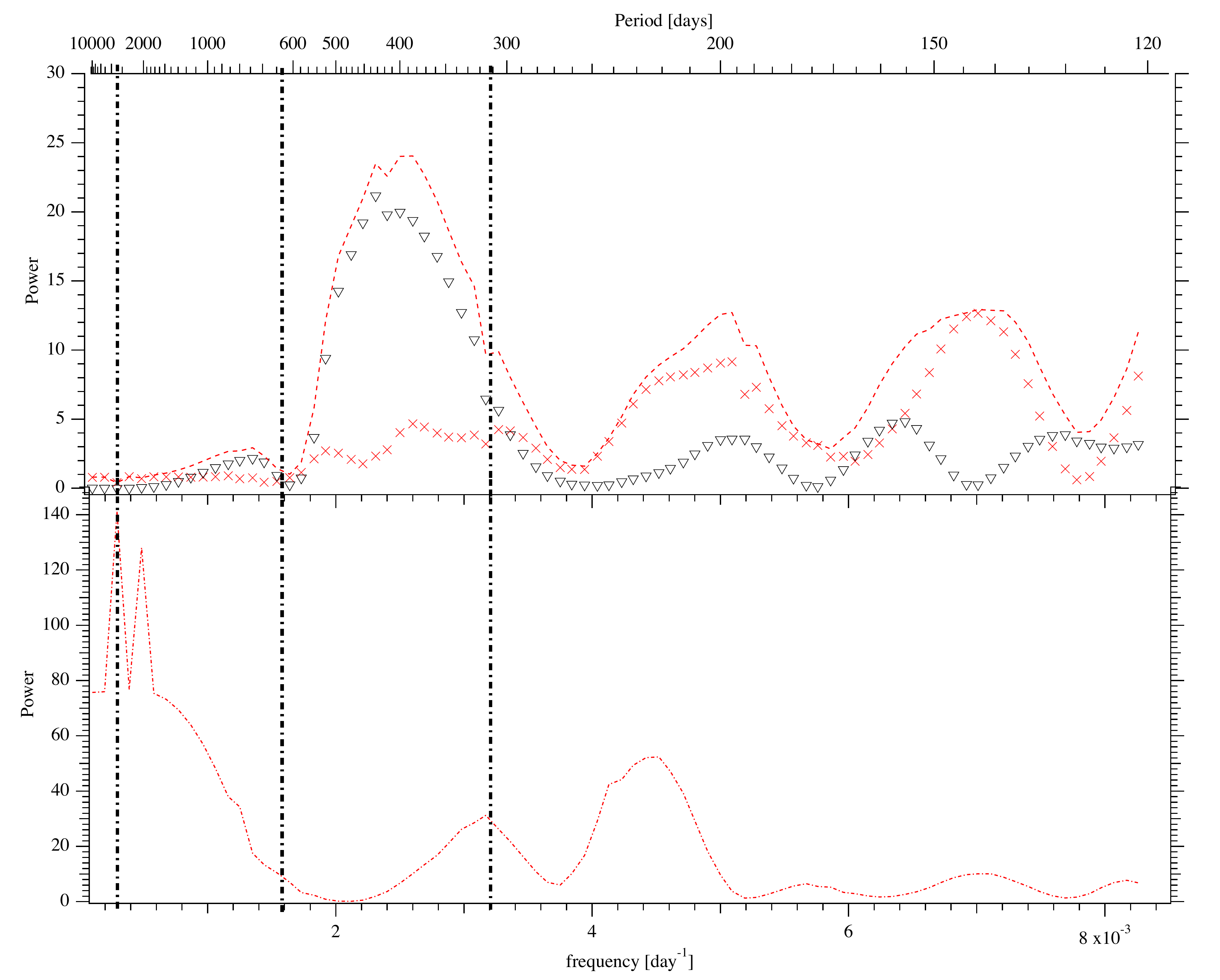}
\caption{Normalized Lomb-Scargle periodograms \citep{Zec09} of: top, \mAs astrometry residuals obtained by applying the Equations 2--5 coefficients to \mA, solving only for \mAs proper motion and parallax;
bottom, window function for the \mAs observation sequence. Vertical lines indicate RV-determined periods for (left to right) components e, b, and d. We find no significant power in the residuals at any component period. } \label{fig-LS}
\end{figure}

Forging ahead, despite the gloomy outlook, we subject those  \mAs astrometric residuals to the following test.  In Figure~\ref{fig-LS} we compare Lomb-Scargle periodograms of astrometric residuals generated before allowing $(O_{n,x})$ and $(O_{n,y})$ to reduce residuals. A periodogram of \mAs residuals to a model without  orbital
  motion (Figure~\ref{fig-LS}, top) contains no significant companion signatures at periods indicated by the RV analysis  (Table~\ref{tbl-RVORB}). 
  
\begin{figure}
\includegraphics[width=5in]{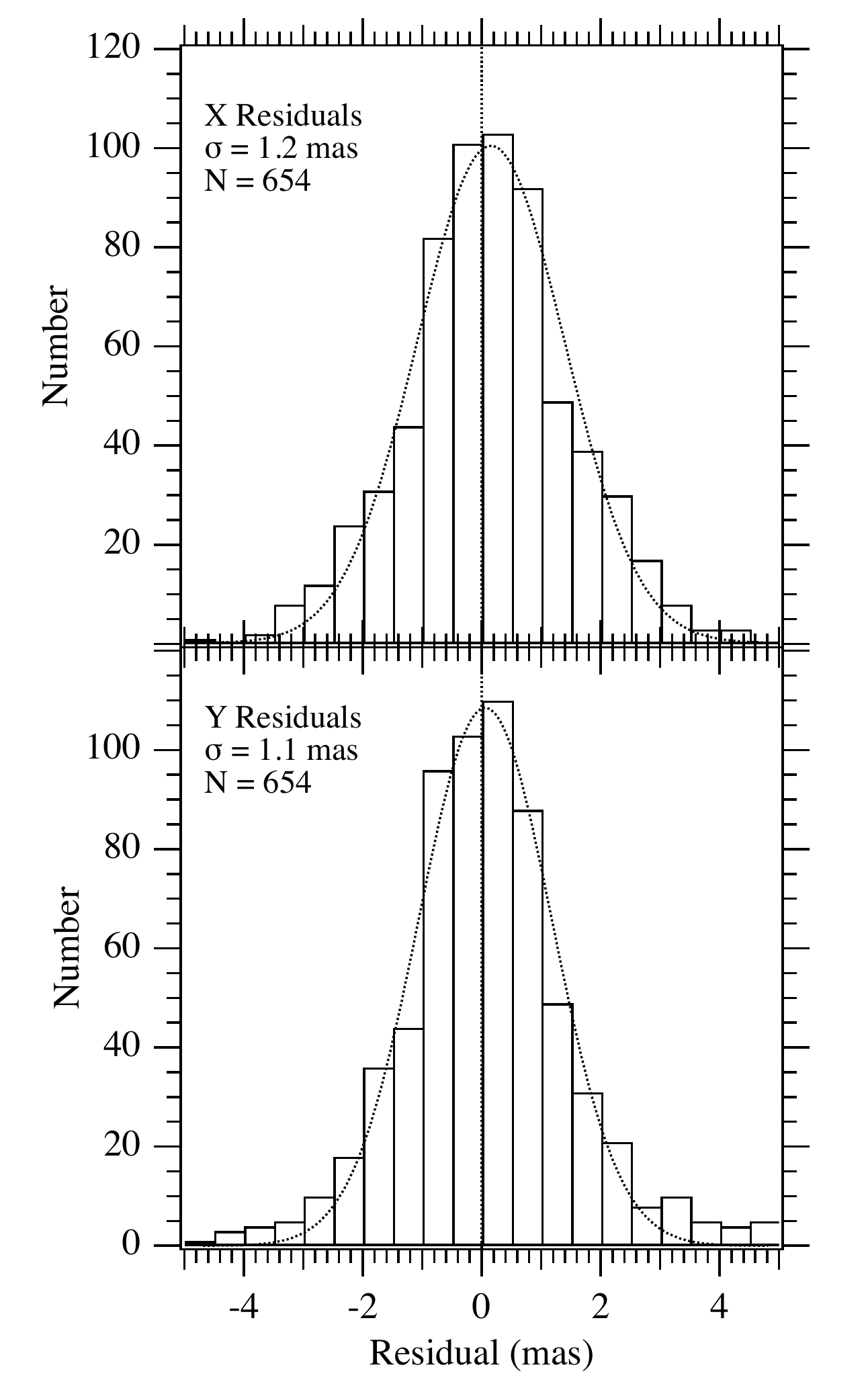}
\caption{Histograms of x and y residuals obtained by deriving the Equation 2-5 coefficients from  654 reference stars
measures,  while modeling reference star parallax and proper motion. The priors for this model had the published EDR3 errors. Distributions are 
fit with gaussians with standard deviations, $\sigma$, indicated in each panel.} \label{fig-FGSH}
\end{figure}

\begin{figure}
\includegraphics[width=6in]{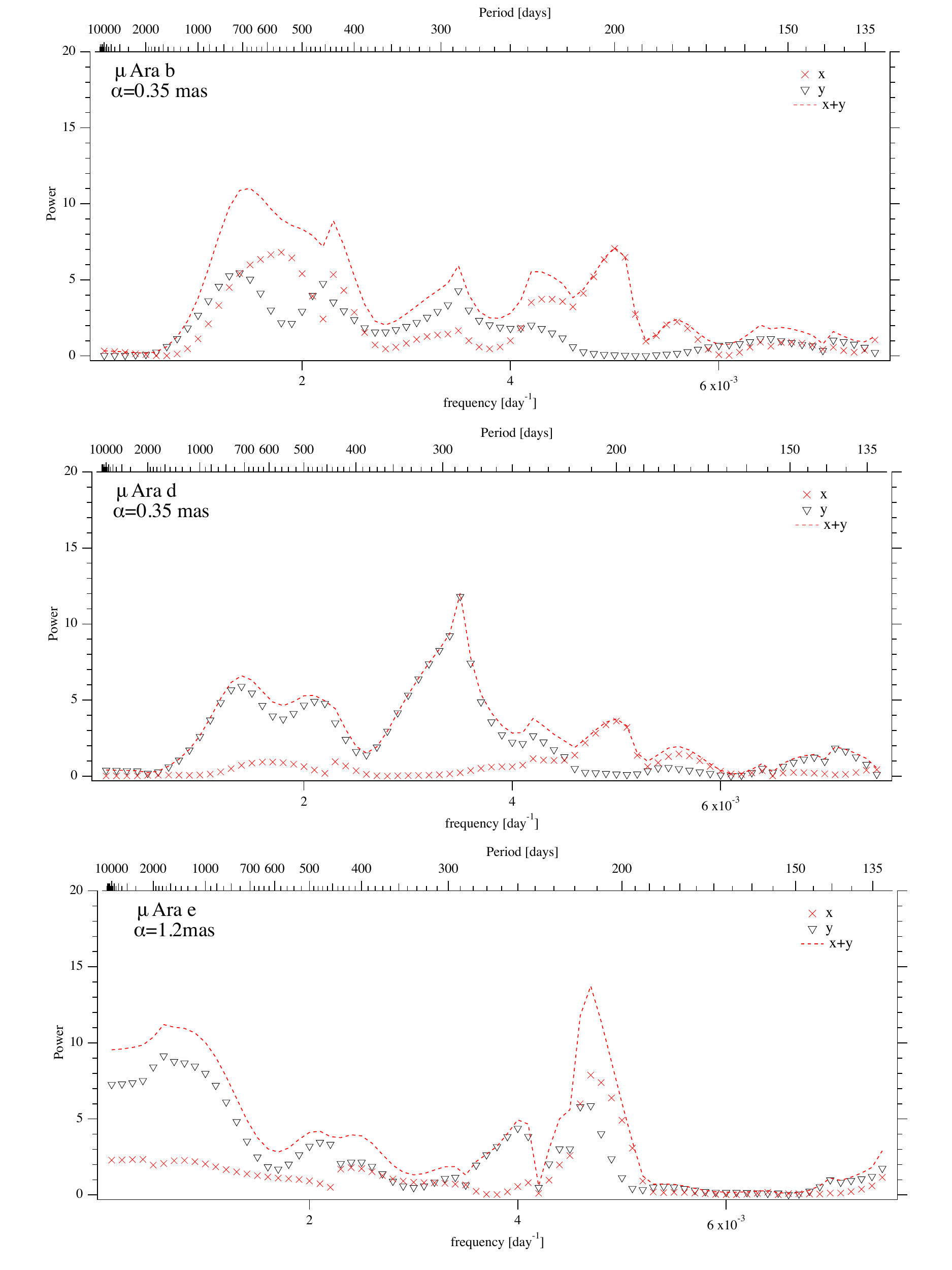}
\caption{Estimated detection thresholds using Lomb-Scargle periodograms for each component, with perturbation amplitude indicated.
A power level of 10 yields a false-positive level of 1\%. } \label{fig-allisLoSt}
\end{figure}

  What could 'hide' in astrometry with per-observation precision a little over 1 mas as demonstarted in Figure~\ref{fig-FGSH}?  We estimate mass upper limits for  the known companions by first populating the \mAs observation dates with Gaussian noise having levels corresponding to the reference star model results in Figure~\ref{fig-FGSH}, $\sigma_{\rm x,y}=1.2, 1.1$ mas. Working with each known companion, \mA\,b, d, e separately, we  add orbital motion, generating signals with various perturbation amplitudes, $\alpha$, using the RV orbital elements from Table~\ref{tbl-RVORB}, holding the unknown longitude of ascending node, $\Omega=0\arcdeg$, and the unknown inclination, $i=0\arcdeg$. For each $\alpha$
  we inspect the periodogram for a signal near the component b,d,e period. An $\alpha$ producing a signal with a false positive level less than 1\% becomes our presumed detection limit, a perturbation we should have seen, given the measured noise level in our astrometry.  We then assume a \mAs mass, 1.13\msune, which, with the known period, provides a companion mass. We provide these mass upper limits in Table~\ref{tbl-F-all} and associated periodograms in Figure~\ref{fig-allisLoSt}. Assuming an expected inclination, $i=60\arcdeg$ (see Section~\ref{BIAS}, below, for the source of this expectation) increases the mass limits by approximately 50\%.


Our measurement precision and extended study duration have  improved the 
accuracy of the parallax of \mA.

\section{Exoplanets with the \FGS} \label{Disc}

Given that \mAs is the final (and only null) result from our originally proposed HST/FGS investigations, we now investigate one aspect of that astrometry.  Our past  exoplanet mass determinations  \citep[table 3][]{Ben17,Ben17b,Ben18b,Ben20b} 
all critically depend
on the inclinations we obtain from our astrometry. These inclinations are listed in Table~\ref{tbl-incs}, along with perturbation semi-major axis, and the two parameters that can signal deviations from a model solving only for proper motion and parallax. 
\subsection{Evidence for Inclination Bias} \label{BIAS}
Table~\ref{tbl-incs} suggests that our exoplanet orbit inclinations seem to skew to small values.
Is there some insidious systematic error in all our analyses that would result in our recovering overly small inclinations, with the result that we find systems to be more face-on than their true orientation?

\begin{figure}
\includegraphics[width=7.5in]{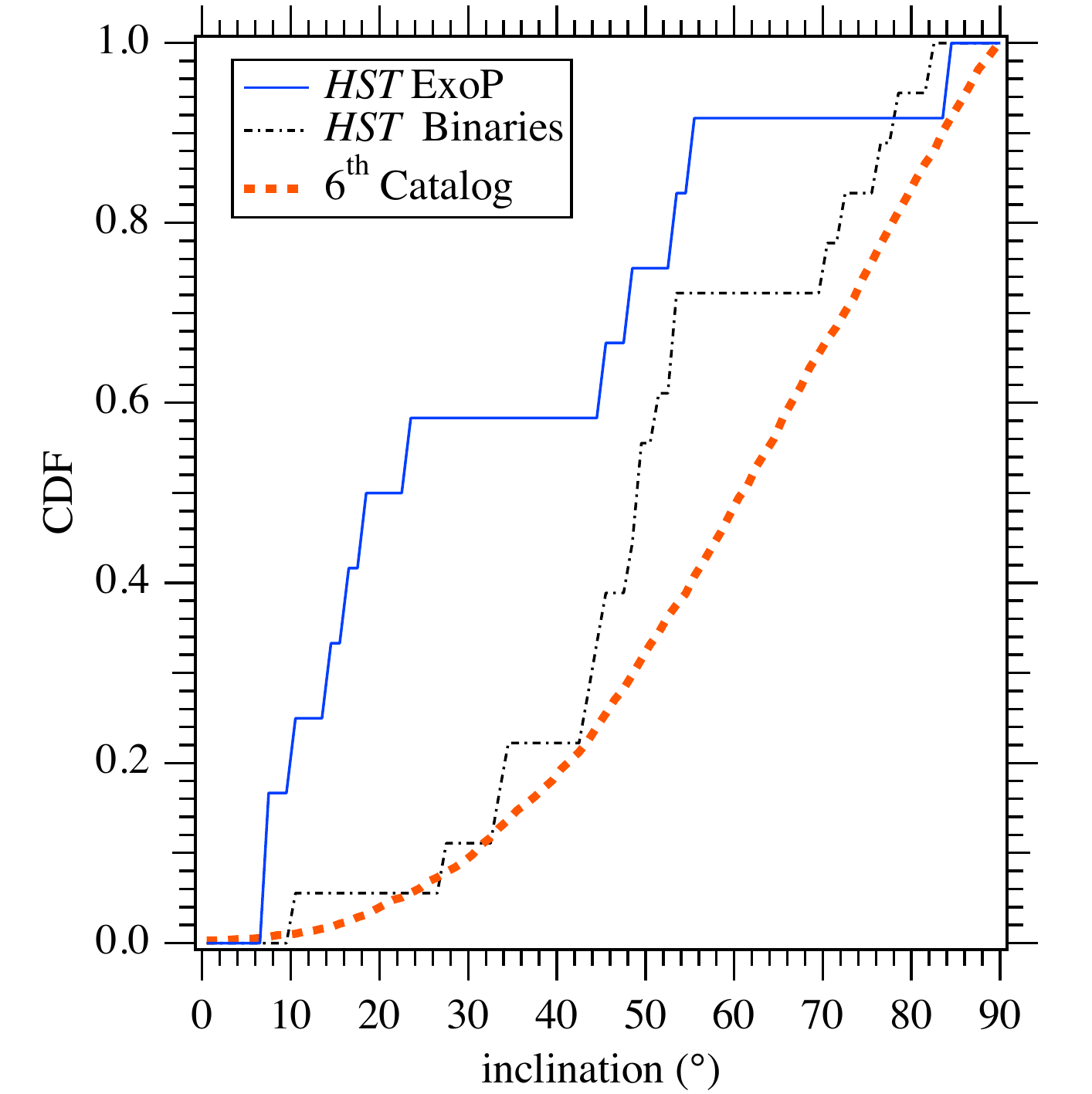}
\caption{ CDFs for; entire inclination set from the 6$^{\rm th}$ Visual Binary Star Catalog  \citep{Har01};  inclinations for \HST-measured binary stars from \cite{Ben16}; and exoplanet perturbation inclinations (Table~\ref{tbl-incs}). KS test results (Table~\ref{tbl-KSres}) indicate  that  neither our exoplanet inclination distributions nor the \HSTs binary distributions are drawn from the same parent population as the 6$^{\rm th}$ Catalog binary inclination population. 
\label{Fig-KStests}}
\end{figure}

We test that our exoplanet perturbation inclinations may not be random by first obtaining a sample of measured inclinations with an assumed random distribution. We harvest over 3200 measured inclinations from The 6$^{\rm th}$ Catalog of Visual Binary Stars \citep[][hereafter WDS]{Har01} and produce the cumulative distribution function (CDF) displayed in Figure~\ref{Fig-KStests}. To produce the CDF we put all inclinations on a 0-90$\arcdeg$ scale by applying this offset to any inclinations over 90$\arcdeg$; $i_{\rm corr}=90\arcdeg-{\rm mod}(i,90\arcdeg)$. A histogram of these inclinations exhibits a peak at $i_{\rm corr}=60\arcdeg$, as expected from  a sample of random orientations. Also plotted are the CDF for the \cite[][table 9]{Ben16} visual binary inclinations (\HSTs Binaries, offset as above), and the CDF for the exoplanetary perturbations listed in Table~\ref{tbl-incs} (ExoP). 

To assess the probability that two CDFs are both drawn from random distributions
we employ a Kolmogorov-Smirnov (KS) test  which produces a test statistic, D, a critical value, C, and a p value, PV. Values of D less than C support the null hypothesis. A p value greater than the adopted significance level (all $\alpha=0.05$) also supports the null hypothesis. 
Table~\ref{tbl-KSres} summarizes the results of KS tests to support or refute the null hypothesis  that   two distributions (first, MLR binary inclinations, then exoplanet inclinations) are drawn from the same parent population (randomly distributed inclinations).

First, because the \cite{Ben16} low-mass Mass-Luminosity Relation (MLR) binary system inclinations do contain a bias towards lower inclinations (those systems being more favorable for discovery, and for the subsequent astrometric measurement required to establish precise stellar masses), as expected, the null hypothesis that  \HSTs MLR inclinations are as random as the 6$^{\rm th}$ Catalog inclinations is not supported, D  marginally larger than C, and the p value lower than the significance level, $\alpha$ (Table~\ref{tbl-KSres}). 
Second, Figure~\ref{Fig-KStests} shows an exoplanet inclination CDF  strikingly dissimilar to the random inclination CDF from the WDS catalog. KS testing  the Table~\ref{tbl-incs} exoplanet inclinations against the known random 6$^{\rm th}$ Catalog inclinations, we find D greatly exceeding C, and p much lower than $\alpha$ (Table~\ref{tbl-KSres}). Our \HSTs astrometrically-derived exoplanet orbit inclinations are clearly inconsistent with a sample with a random distribution of inclinations.

\subsection{Possible Bias Explanations}
\begin{figure}
\includegraphics[width=6in]{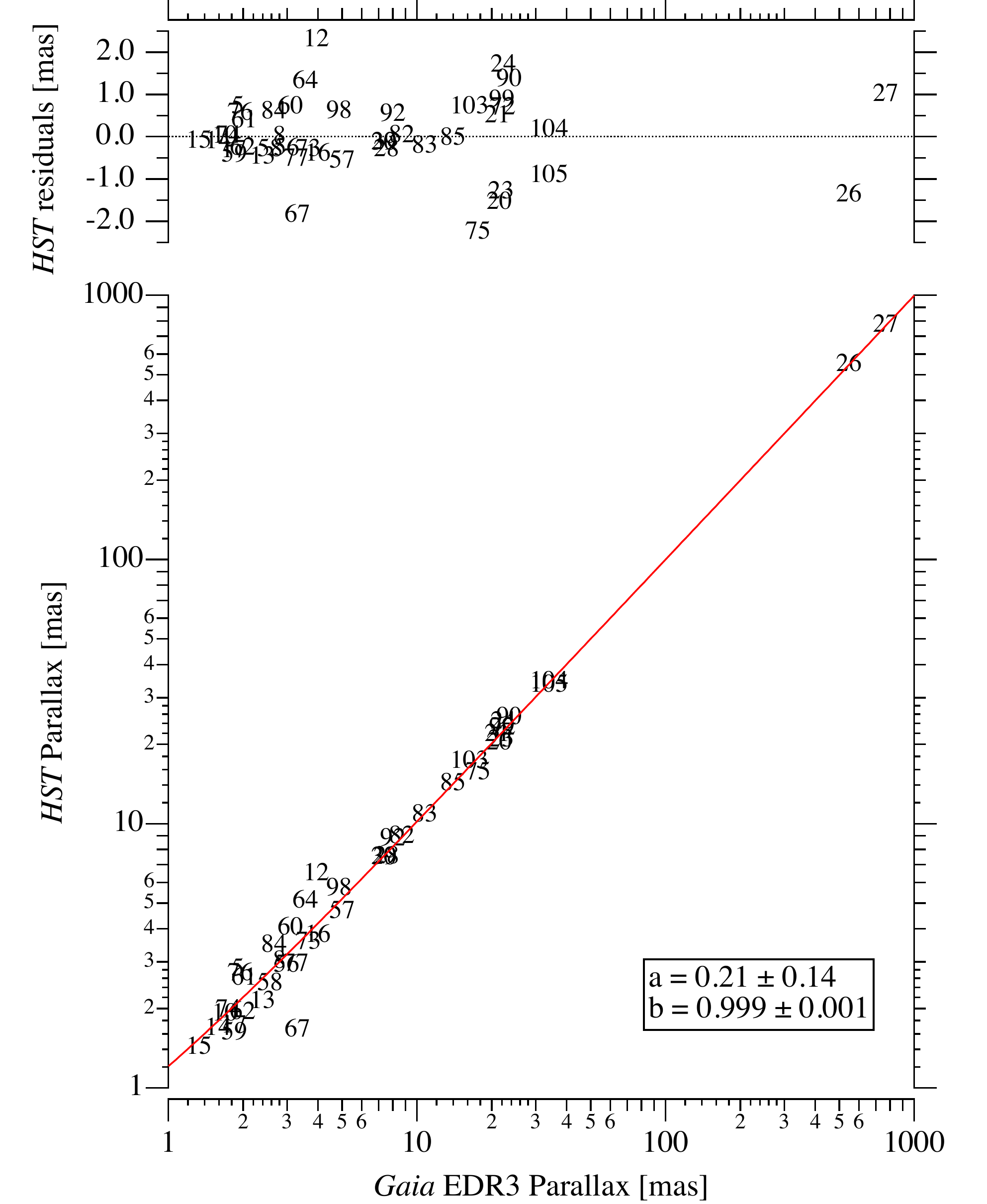}
\caption{ \HST-FGS parallaxes compared with \Gs EDR3 parallaxes. The linear regression assumes no errors for \G. 
Comparison plots only targets with \Gs $RUWE<1.4$, ID numbers from Benedict et al. (2017), table 1: 5, 8, 10, 12, 13, 14, 15, 16, 17, 20, 21, 23, 24, 26, 27, 28, 29, 30, 56, 57, 58, 59, 60, 61, 62, 64, 67, 72, 73, 74, 75, 76, 77, 82, 83, 84, 85, 90, 92, 98, 99, 103, 104, and 105. The \HSTs residuals have  an rms value 0.82 mas.
\label{Fig-GH}}
\end{figure}

We now explore four potential areas which could produce the observed bias in our exoplanet system inclinations; small number statistics, modeling technique, FGS mechanical issues, and
orbit modeling of noise-dominated data. None adequately explains the clearly demonstrated bias.
\begin{enumerate}
\item Small Number of Inclinations

To explore any possible effect of comparing unequal sampling sizes, we  drew from the 6$^{\rm th}$ Catalog 100,000  randomly selected samples of inclination, 
each of 12 values representing the exoplanet sample. Running KS tests comparing each sample CDF with the 6$^{\rm th}$ Catalog  CDF we find only a 5\% probability of a randomly selected sample CDF disagreeing with the 6$^{\rm th}$ Catalog. 
This Monte Carlo test suggest that small number statistics are highly unlikely to be the cause of the  exoplanet inclination bias.

\item Restricted Modeling

\HIP~intermediate astrometric data (IAD) have been used in several studies to estimate the mass or upper mass limits for possible exoplanets \citep[e.g.][]{Maz99, Ref11}. \cite{Pou01} found that some small inclinations  were merely artifacts of the fitting procedure that was used. Fitting ($i, \Omega$) to the HIP IAD, where the $\alpha\,{\rm sin}\,i$ is \underline{much smaller} than the astrometric precision, always yields low
values of sin\,$i$, regardless of the true inclination. 

For our analysis we force astrometry and RV to describe the same perturbations
through this constraint \citep[e.g.][]{Pou00}, shown for a perturbing companion b. 
\begin{equation}
\displaystyle{{\alpha_{\rm b}~sin\,i_{\rm b} \over \varpi_{abs}} = {P_{\rm b} K_{\rm b} (1 -
e_{\rm b}^2)^{1/2}\over2\pi\times4.7405}} 
\label{PJ}
\end{equation}
\noindent Equation~\ref{PJ} contains quantities derived from astrometry (parallax, $\varpi_{abs}$, 
host star perturbation orbit size, $\alpha$, and inclination, $i$) on the left hand side (LHS), and 
quantities  derivable from both (the period, $P$ and eccentricity, $\epsilon$), or only radial 
velocities  (the RV amplitude of the primary, $K$, induced by a companion), on the right hand side (RHS). \HST~time is in high demand. This, in most cases results in sparse orbit coverage of any  perturbation afforded by the astrometry. Therefore,
the RV data were always essential in determining a perturbation orbit. For a multi-component system, n=1,2,4 (for example, \mAs b,d,e), $O_{n,x}$ and $O_{n,y}$ in Equations 3 and 4 are functions of the classic orbit parameters. They describe the  motion (on the sky and in RV) of the parent star around the barycenter.
The RV cover a far greater time span for each component perturbation, providing essential
support for determining $P$, $\epsilon$, $K$, $\omega$, and $T_0$.

While for our analysis we do use a relationship between the astrometry and the RV (see Equation ~\ref{PJ}), our
modeling is significantly different than the modeling of the HIP IAD. We hold no orbital or astrometric parameters as constants. Our solutions do not converge unless there is a measurable signal.
Given the relatively short time span for the astrometric measures, our past and present analyses critically depend on both
 RV measures secured over longer time spans and the Equation~\ref{PJ} relation between astrometry (LHS) and RV (RHS).   For most of the targets in Table~\ref{tbl-incs} the  period, amplitude, and eccentricities from RV-only are well determined, with errors insufficient to much change the LHS inclination via Equation~\ref{PJ}. To increase the inclination requires a decrease in either parallax or perturbation size, or both. 
 
\item FGS at Fault

 We now estimate possible errors for our parallax and perturbations, using  HD\,202206\,c as a test case \citep{Ben17b}. Figure~\ref{Fig-GH}  compares a subset of FGS parallaxes (Benedict et al. 2017, table 1) with Gaia EDR3 \citep{Lin21a}, where the subset satisfies \Gs $RUWE<1.4$.  Based on the EDR3 error assessments of \cite{Sta21} and \cite{Lin21b}, we assume that \Gs parallaxes are error-free, a reasonable assumption given the $\sim0.03$ mas errors compared to an average $\sim0.19$ mas errors for \HST. Figure~\ref{Fig-GH}  yields FGS parallax errors typically less than 1 mas, assuming \Gs perfection. Holding all other terms in Equation~\ref{PJ} constant, to increase the astrometric inclination for HD\,202206\,c from the measured $7\arcdeg$ to, for example, $40\arcdeg$ would require a parallax, $\varpi=105$ mas \citep[the][value is 21.96 mas]{Ben17b}, a parallax mis-measurement far exceeding what we have achieved in the past. This leaves only the perturbation size, $\alpha$, suspect. 

All FGS parallax measurements have built-in constraints similar to those we employ to derive an exoplanet perturbation, $\alpha$. The precisely known period and eccentricity of the orbit of the Earth serve as the RHS terms of Equation~\ref{PJ}. The calculated parallax factors encode those terms plus a perceived inclination, basically the ecliptic latitude of the parallax target.  As demonstrated in Figure~\ref{Fig-GH}, our errors in determining the parallactic ellipse size rarely exceed 1 mas. The majority of those results came from campaigns with N=9-11 measurement sets \citep[e.g.][]{Ben07}. Astrometric accuracy scales as $1/\sqrt{N}$. Thus, with 31 observational epochs we might realistically expect  errors in HD\,202206 $\alpha$ and $\varpi$ of $\sim0.5$ mas. Increasing the inclination to $40\arcdeg$ would decrease the HD\,202206 perturbation to $\alpha=0.16$ mas. To yield the average inclination expected by the assumption that orbit angular momentum 
vectors for exoplanet systems are randomly and isotropically distributed ($60\arcdeg$) would require $\alpha=0.12$ mas, an unlikely $6\sigma$ difference. 


Our Monte-Carlo tests show how unlikely that we, by chance, studied many exoplanetary systems with lower than expected inclinations. However, two recent results argue for small-number statistics rather than systematic bias. These systems also yield low inclinations. The first is \citep{Ben20b} 
Proxima Centauri c, with an inclination, $i_{\rm c}=18\pm4\arcdeg$, which modulo $90\arcdeg$, agrees with \cite{Ker20}. The second is vA 351, a complex binary in the Hyades, consisting of components AD and BC with a 2.7y orbital period, and components BC in a 0.75 day orbital period  \citep{Ben21}. FGS fringe tracking, fringe scanning, and independent speckle camera observations yield an AD-BC inclination, $i=14\pm8\arcdeg$.
Extensive RV measurements yield a mass ratio for components B/C. That ratio, coupled with a total BC mass from FGS astrometry, yield masses for the B and C components which agree within 7\% of those predicted from the \cite{Ben16} Mass-Luminosity relation, further confirming the validity of our measured system inclination.

\item Fitting an Orbit to Astrometric Noise

For this test we choose $\kappa$ Pavonis, a dwarf Cepheid, previously a parallax target \citep[][parallax $\varpi=5.57\pm0.28$ mas]{Ben11}. The modeling resulted in a $\chi^2/{\rm DOF}=0.426$ and an rms residual of 1.9 mas. This parallax agrees within the errors with the \Gs EDR3 value,
$\varpi=5.24\pm0.12$ mas. For $\kappa$ Pav, $RUWE=2.29$, a high value likely due to photometric variability and brightness ($G<6$), there yet being no astrometric, RV, or direct imaging evidence of a companion.  The \cite{Bra21} $\chi^2$=6.58, is close to the median value, 5.4, found for a random sample of similarly bright stars (Section~\ref{nada}). We modified the model to solve for an orbit, including totally fictitious priors for period, $P=435\pm3^{\rm d}$, eccentricity, $e=0.3\pm0.1$, time of periastron passage, $T_0=53041\pm30^{\rm d}$, longitude of periastron passage, $\omega=269\arcdeg \pm17\arcdeg$, and RV amplitude, $K=113\pm20$ \ms. 
This produces an orbit with a perturbation size, $\alpha=0.4\pm0.2$ mas and an inclination, $i=2.9\arcdeg \pm1.5\arcdeg$. The $\chi^2/{\rm DOF}=0.417$ is only 2\% less than a model without the orbit. The rms residual is unchanged at 1.9 mas. While including an orbit did produce a result with a very small decrease in $\chi^2/{\rm DOF}$, we find  no effect on the rms residual. That, and the very low statistical significance of the $\alpha$ and inclination values (2$\sigma$), demonstrate a companion non-detection. Our previous inclination and perturbation results (Table~\ref{tbl-incs}) are all $>5\sigma$, demonstrably not a result of fitting noise.
\end{enumerate}

\section{System Stability} \label{Disc1}
\mAs has always presented stability challenges \citep{Pep07, Tim13, Las17,Agn18}. Our remodeling of a larger set of RV supports the conclusion of \cite{Tim13} that \mAs b and d are $near$ a 2:1 resonance; $P_{\br}/P_{\dr} = 2.095\pm0.002$. Our improved period for the outermost companion, e,  places it $near$ a 6:1 resonance with component b, $P_{\er}/P_{\br} = 6.12\pm0.04$. Note  that \cite{Las17} include \mAs (and the Solar System) among the unstable systems.

However, our incomplete characterization of the \mAs system (minimum masses from Section~\ref{nada}) fails to provide a solution to the vexing problem of stability.
The orbital periods are similar to those of Earth, Mars, and Jupiter, but of course
the masses are much larger. These
features suggest that gravitational interactions should induce large-amplitude
oscillations in the orbital elements, potentially resulting in ejections or
collisions between the orbiting bodies. Here we examine these
interactions with analytic and N-body methods. We find the results inferred from
the astrometric and radial velocity observations predict an unstable system.

We first consider the Hill stability \citep{SzebehelyZare77,MarchalBozis82,Gladman93} of
the 3 planet-planet pairs to assess the likelihood of orbital stability. Hill stability
is only strictly applicable to a three-body system outside of resonance, but it is
analytic and can provide an approximate assessment of stability in more complicated
systems such as $\mu$ Arae. Following the prescription of \cite{BG06_stab,BG07_res},
we characterize the Hill stability via the ratio $\beta/\beta_{crit}$ in which ratios
less than 1 indicate instability and ratios greater than 1, stability. We use the publicly
available code \HillStability\footnote{https://github.com/RoryBarnes/HillStability}
to calculate this value and find that the b-d pair is at the limit with $\beta/\beta_{crit} = 1.0$.
The other pairs appear comfortably stable.

We support this predictions with a direct N-body simulation. We integrated our best fit system
with the \spinbody module in
\vplanet~\citep{Bar20}\footnote{\vplanet is publicly available at
https://github.com/VirtualPlanetaryLaboratory/vplanet} to model evolution from first principles.
We used a fixed timestep
of 0.365 days corresponding to nearly 1000 steps per orbit for planet d with \vplanet's
4th order Runge-Kutta scheme, which is generally small enough to capture the evolution.
We find that the system breaks apart due to interactions between planets b and d in less than $10^5$ years, confirming the instability
predicted by the Hill theory.

Thus, \mAs continues to be problematic in terms of
orbital stability, but our Hill stability analysis suggests that modest changes
to the system, particularly the masses and orbits of planets b and d, could
result in a stable system. Alternatively, we find that if we remove planet d from
the system, then the resulting orbital evolution is regular and long-term stability appears likely.

\section{Summary} \label{Summ}

For  the \mAs system from a model which utilizes {\it HST/FGS} astrometry and ground-based RV we find;
\begin{enumerate}
\item significantly improved companion orbital elements ($P$, $\epsilon$, $\omega$, $T_0$, $K$), derived from only the large body of RV data,

\item with a model containing no proper motion and parallax priors  for \mAs a parallax, $\pi_{abs}=64.11\pm0.13$ mas, agreeing with the \HIP~and $Gaia$ EDR3 values within the errors, and a  proper motion relative to a \Gs EDR3 reference frame, $\vec{\mu} = 190.83$ mas  yr$^{-1}$
with a position angle, P.A. = $184\fdg3$, differing by +0.66 mas yr$^{-1}$ and $-0\fdg2$ compared to \Gs EDR3,

\item that astrometric residuals  of order 1 mas to models solving only for parallax and proper motion  contain no evidence for any of the known companions of \mA,

\item assuming those levels of measurement precision yields lower limits for \mA\, b, d, e of 4.3\mjupe, 7.0\mjupe, and 4.4 \mjupe,

\item  that KS testing supports the assertion that exoplanetary orbit inclinations previously measured with the \HSTs FGS are  biased towards small inclinations. Based on comparisons with \Gs EDR3 parallaxes, the results from an orbit determination when none exists, and independently confirmed recent results, we argue that this could be chance, not systematic error.

\item an inherently unstable system, if it includes \mAs \dr

\item a system stable for 10$^6$ y without \mAs \dr,

\end{enumerate}

Finally, all \HSTs FGS exoplanet results represent a useful test of \Gs results. With  10--100 {\it micro}second of arc precision, and a longer time span for astrometric observations, \Gs will certainly improve on those results,  either exposing a bias in FGS exoplanet astrometry, or not. If  \HSTs FGS exoplanet results do contain a bias, then \Gs investigators, who  will produce a large number of perturbation orbital elements with perturbations near the \Gs per-observation precision, should be aware of this possibility.
We hope that a future combination of \FGSs and RV data with \Gs can improve the accuracy of any astrometric result, and definitively produce companion orbits and masses.


\begin{acknowledgments}

We thank Dr. Tim Brandt for his careful, well-reasoned referee's report which significantly improved  the paper content, and an initial anonymous referee for flag raising. This work is based on observations made with the NASA/ESA Hubble Space Telescope, obtained at the Space Telescope Science Institute. Support for this work was provided by NASA through grants 11210 and 11788
 from the Space Telescope 
Science Institute, which is operated
by the Association of Universities for Research in Astronomy, Inc., under
NASA contract NAS5-26555. RB acknowledges support from the NASA Astrobiology Program Grant Number 80NSSC18K0829. This publication makes use of data products from the 
Two Micron All Sky Survey, which is a joint project of the University of 
Massachusetts 
and the Infrared Processing and Analysis Center/California Institute of 
Technology, 
funded by NASA and the NSF.  This research has made use of the {\it SIMBAD} and {\it Vizier} databases, 
operated at Centre Donnees Stellaires, Strasbourg, France; Aladin, developed and maintained at CDS; the NASA/IPAC Extragalactic Database (NED) 
which is operated by JPL, California Institute of Technology, under contract 
with 
NASA;  and NASA's truly essential Astrophysics Data System Abstract Service. This work has made use of data from the European Space Agency (ESA)
mission {\it Gaia} (\url{http://www.cosmos.esa.int/gaia}), processed by
the {\it Gaia} Data Processing and Analysis Consortium (DPAC,
\url{http://www.cosmos.esa.int/web/gaia/dpac/consortium}). Funding
for the DPAC has been provided by national institutions, in particular
the institutions participating in the {\it Gaia} Multilateral Agreement. Many people over the years have materially improved all aspects of the work reported, particularly Linda Abramowicz-Reed, Art Bradley, Denise Taylor, and all the co-authors of our many papers. 
We thank Dr. Paul Robertson for use of his code based on the Zechmeister and K\"{u}rster normalized Lomb-Scargle algorithm.
The American Astronomical Society supported the preparation of this paper 
while GFB carried out his duties as Society Secretary. For this he is sincerely grateful.
GFB fondly remembers Debbie Winegarten (R.I.P.), whose able assistance with Secretarial matters freed him to devote  time to this analysis. 
\end{acknowledgments}

\bibliography{/Active/myMaster}

\clearpage

\begin{deluxetable}{l l r}
\tablewidth{0in}
\tablecaption{\mAs Stellar Parameters \label{tbl-STAR}}
\tablehead{ 
\colhead{Parameter}& 
\colhead{Value}& 
\colhead{Source\tablenotemark{a}}
}
\startdata
SpT&G3IV-V&1\\
T$_{eff}$&5773 K&6\\
log g&4.2 $\pm$ 0.1&6\\
$[Fe/H]$&0.28 $\pm$ 0.03&6\\
age&5.7 $\pm$ 0.6 Gy&5\\
mass&1.13 $\pm$ 0.02 $\cal{M}_{\sun}$&5\\
distance&15.57 $\pm$ 0.02 pc & 2\\
Radius&1.33 $\pm$ 0.02$ R_{\sun}$&5\\
v\,sin\,i&3.1 $\pm$ 0.5 \kms&7\\
m-M&0.961$\pm$ 0.005&2\\
$V$&5.15 $\pm$ 0.01&1\\
$K$&3.68 $\pm$ 0.25&3\\
$V-K$&1.47 $\pm$ 0.25&1,3\\
\enddata
\tablenotetext{a}{1 = SIMBAD,
2 = this paper,\\ 3 = 2MASS, 5 = \cite{Bon15},\\ 6 = \cite{Sot18}
7 = \cite{Fis05}.}
\end{deluxetable}


\begin{deluxetable}{l l l l l}
\tablecaption{Radial Velocities\tablenotemark{a} \label{tbl-newRV}}
\tablewidth{0in}
\tablehead{ 
\colhead{mJD\tablenotemark{b}}&
\colhead{RV} &
\colhead{RVerr}&
\colhead{residual} &
\colhead{source\tablenotemark{c}}
}
\startdata
52906.5194&-9.29090&0.00118&0.00516&11\\
53160.7260&-9.33980&0.00070&0.00105&11\\
53161.7278&-9.34280&0.00070&0.00017&11\\
53162.7260&-9.34480&0.00070&0.00038&11\\
53163.7259&-9.34770&0.00070&-0.00130&11\\
53164.7258&-9.34820&0.00070&-0.00220&11\\
53165.6828&-9.34550&0.00070&-0.00086&11\\
53166.7820&-9.34270&0.00070&0.00036&11\\
53167.7269&-9.34210&0.00070&0.00018&11\\
53201.6199&-9.36110&0.00119&-0.00091&11\\
53202.6414&-9.35980&0.00119&0.00114&11\\
53203.6108&-9.36190&0.00119&-0.00175&11\\
...&...&...&...&...
\enddata
\tablenotetext{a}{Full table available electronically. All velocity units in \kms.}
\tablenotetext{b}{mJD=JD-2400000}
\tablenotetext{c}{11 = HARPS1 \citep{Pep07}, 12 = CORALIE \citep{Pep07}, 14 = AAT \citep{Tin01, Wit14, Wit17} and this paper,  
15 = HARPS2  \citep{Loc15}}
\end{deluxetable}

\begin{deluxetable}{c r}
\tablecaption{RV Offsets \label{tbl-gammas}} 
\tablewidth{6in}
\tablehead{
\colhead{RV Source} &
\colhead{$\gamma$ [\ms]} 
		}
\startdata
CORALIE&-9379.1$\pm$0.9\\
HARPS1&-9348.1 0.2\\
AAT&-7.6 0.3\\
HARPS2&1.7 0.2
\enddata
\end{deluxetable}

\begin{deluxetable}{c  r r r r}
\tablecaption{Orbital Elements for the \mAs b, c, d, e Perturbations, Radial Velocity Only\label{tbl-RVORB}}
\tablewidth{6in}
\tablehead{
\colhead{Parameter} &  
\colhead{b} &
\colhead{c} &
\colhead{d} &
\colhead {e} 
		}
\startdata
P [days]&645.0$\pm$0.3&9.6392$\pm$0.0006&307.9$\pm$0.3&3947$\pm$23\\
P [yrs]&1.7664 0.0008&0.026391 0.000002&0.8429 0.0008&10.81 0.06\\
T [mJD]&52396 28&52 4&52720 9&53264 388\\
$\epsilon$&0.036 0.007&0.16 0.06&0.091 0.014&0.022 0.012\\
K [m/s]&36.1 0.2&2.94 0.17&12.23 0.27&22.18 0.25\\
$\omega$ [$\arcdeg$] &39 16&197 20&193 10&84 36
\enddata
\end{deluxetable}

\begin{deluxetable}{l l l r r r r r r r r r}
\tabletypesize{\tiny}
\tablewidth{0in}
\tablecaption{\mAs Field Astrometry\tablenotemark{a} \label{tbl-DATA}}
\tablehead{ 
\colhead{Set}& 
\colhead{Star}& 
\colhead{\HST ID} &
\colhead{V3 roll} &
\colhead{X} &
\colhead{Y} &
\colhead{$\sigma_X$} &
\colhead{$\sigma_Y$} &
\colhead{t$_{\rm obs}$} &
\colhead{P$_{\alpha}$} &
\colhead{P$_{\delta}$}
}
\startdata
1&3&F9YM3703M&143.36&-6.35877&2.36891&0.00162&0.00179&54289.5332&-0.547823&-0.446584\\
1&3&F9YM370DM&143.36&-6.35864&2.36810&0.00164&0.00195&54289.5430&-0.548134&-0.446536\\
1&3&F9YM3709M&143.36&-6.35758&2.37029&0.00157&0.00182&54289.5390&-0.548009&-0.446558\\
1&3&F9YM370KM&143.36&-6.35756&2.36914&0.00154&0.00174&54289.5505&-0.548353&-0.446493\\
1&20&F9YM3707M&143.36&60.02043&75.67643&0.00203&0.00224&54289.5375&-0.547843&-0.446125\\
1&20&F9YM370CM&143.36&60.02464&75.67678&0.00210&0.00230&54289.5422&-0.547992&-0.446101\\
1&21&F9YM370FM&143.36&-58.68031&99.86974&0.00209&0.00172&54289.5446&-0.549343&-0.446231\\
1&21&F9YM3705M&143.36&-58.67832&99.87097&0.00197&0.00211&54289.5356&-0.549057&-0.446278\\
1&21&F9YM370AM&143.36&-58.67787&99.87108&0.00195&0.00202&54289.5399&-0.549194&-0.446257\\
1&22&F9YM3708M&143.36&56.18870&40.94541&0.00218&0.00198&54289.5383&-0.547666&-0.446275\\
1&22&F9YM3702M&143.36&56.18970&40.94592&0.00234&0.00209&54289.5324&-0.547478&-0.446300\\
1&22&F9YM370LM&143.36&56.18984&40.94558&0.00241&0.00238&54289.5513&-0.548056&-0.446201\\
1&22&F9YM370EM&143.36&56.19062&40.94591&0.00213&0.00205&54289.5437&-0.547839&-0.446245\\
1&26&F9YM370IM&143.36&-30.61732&-95.19696&0.00247&0.00302&54289.5482&-0.547849&-0.446956\\
1&26&F9YM3704M&143.36&-30.61421&-95.19866&0.00279&0.00284&54289.5344&-0.547419&-0.447030\\
1&27&F9YM370JM&143.36&113.50825&-60.51056&0.00223&0.00225&54289.5495&-0.546776&-0.446515\\
2&3&F9YM3809M&148.22&-5.44870&2.63273&0.00193&0.00191&54293.4618&-0.647938&-0.433618\\
2&3&F9YM380PM&148.22&-5.44869&2.63362&0.00184&0.00183&54293.4778&-0.648432&-0.433538\\
...&...&...&...&...&...&...&...&...&...&...&...
\enddata
\tablenotetext{a}{Set (orbit) number, star number (\#3 = \mA; reference star numbers same as Table~\ref{tbl-piref}), \HSTs orbit and target identifier, spacecraft +V3 axis roll angle as defined in Chapter 2, \FGSs Instrument 
Handbook \citep{Nel15a}, OFAD-corrected X and Y positions in arcsec, position measurement errors in arcsec, time of observation = JD - 2400000.5, RA and DEC parallax factors. We provide a complete table in the
electronic version of this paper.}
\end{deluxetable}


\begin{deluxetable}{l r r r r r r r}
\tablewidth{0in}
\tablecaption{Parallax Priors for \mAs Astrometric Reference Stars\label{tbl-piref}}
\tablehead{\colhead{Ref Star \#}
 & \colhead{$V$\tablenotemark{a}}
  & \colhead{$B-V$}
  & \colhead{SpT\tablenotemark{b}}
  &  \colhead{ EDR3 Source }
 & \colhead{$G$}
&  \colhead{$\varpi$ \tablenotemark{c}}
& \colhead{$RUWE$ \tablenotemark{d}}
 		}
\startdata
20&12.22&1.50$\pm$0.03&     K3III&5946035772071080000&11.6655&0.48$\pm$0.02&0.981\\
21&12.11&1.09 0.03&    G8III&5945942146094740000&11.7453&0.62 0.01&0.976\\
22&12.96&0.61 0.03&    F6V&5946035776383010000&12.7726&1.68 0.01&0.887\\
24&14.79&1.37 0.07&K1III&5945941974282210000&14.370&0.01 0.04&1.968\\
26&15.27&0.87 0.08&     K1V&5945930154546000000&14.9807&1.32 0.03&1.028\\
27&14.69&0.54 0.06&     F4V&5946024059712400000&14.5239&0.57 0.02&0.941
\enddata
\tablenotetext{a}{V, B-V ex SMARTS 0.9m \citep{Sub10}}
\tablenotetext{b}{Spectra obtained with the RC Spectrograph on the CTIO Blanco 4m.}
\tablenotetext{c}{Parallax in mas with EDR3 errors. Modeling used uniform 1 mas error for all priors.}
\tablenotetext{d}{Reduced unit weight error from EDR3}
\end{deluxetable}

\begin{deluxetable}{c r r r r}
\tablewidth{0in}
\tablecaption{Reference Star Relative Positions\tablenotemark{a} and Measured Parallax\tablenotemark{b}   \label{tbl-POS}}
\tablehead{
\colhead{Star}&
 \colhead{$\xi$ } &
   \colhead{$\eta$} &
   \colhead{$\varpi$} &
   \colhead{$RUWE$}
   		}
\startdata
20&-24.80913$\pm$0.00015&100.22019$\pm$0.00014&0.28$\pm$0.17&0.981\\
21&-115.48324 0.00010&19.89530 0.00010&1.39 0.13&0.976\\
22&0.65176 0.00012&76.26965 0.00010&1.73 0.13&0.887\\
24&-66.90436 0.00021&18.47734 0.00018&-0.77 0.23&1.968\\
26&57.26613 0.00021&-74.94361 0.00020&1.28 0.21&1.028\\
27&116.18275 0.00018&61.11125 0.00016&0.76 0.19&0.941
\enddata
\tablenotetext{a}{~Units are arc seconds, rolled to RA ($\xi$) and DEC ($\eta$), epoch 2008.6524 (J2000). Roll uncertainty $\pm0\fdg02$.}
\tablenotetext{b}{Final values from a model with input parallax prior errors 1 mas and input proper motion priors 1 mas yr$^{-1}$.}
\tablenotetext{c}{RA = 266.0392504	, DEC = -51.8140955, J2000}

\end{deluxetable}

\begin{deluxetable}{c r r r r}
\tablewidth{0in}
\tablecaption{Reference Star Proper Motions\tablenotemark{a} and Agreement with EDR3\tablenotemark{b}   \label{tbl-mus}}
\tablehead{
\colhead{Star \#}&
 \colhead{$\mu_{\rm RA}$ } &
   \colhead{$\mu_{\rm Dec}$} &
   \colhead{$\Delta \mu_{\rm RA}$} &
   \colhead{$\Delta \mu_{\rm Dec}$}
   		}
\startdata
20&-3.06$\pm$0.19&-2.91$\pm$0.16&-0.53&0.14\\
21&2.38 0.12&0.21 0.12&0.02&0.01\\
22&0.79 0.14&-16.63 0.12&0.42&-0.33\\
24&-5.31 0.24&-5.85 0.21&0.11&0.06\\
26&-18.52 0.22&-17.55 0.20&-0.11&-0.02\\
27&1.99 0.21&-2.88 0.18&0.06&0.08
\enddata
\tablenotetext{a}{~Units: mas yr$^{-1}$}
\tablenotetext{b}{Difference from a model with input parallax prior errors 1 mas and input proper motion priors 1 mas yr$^{-1}$.}

\end{deluxetable}


\begin{deluxetable}{ll}
\tablecaption{Reference Frame Statistics, \mAs Parallax, and Proper Motion\label{tbl-SUM}}
\tablewidth{0in}
\tablehead{\colhead{Parameter} &  \colhead{Value} }
\startdata
Study duration  &2.85 y  \\
number of observation sets    &   32  \\
reference star $\langle V\rangle$ &  13.67     \\
reference star $\langle (B-V) \rangle$ &1.00   \\
{\it HST}: model with reference star EDR3 prior errors\\ 
~~~~~~~Absolute $\varpi$& 63.84 $\pm$ 0.13    mas \\
~~~~~~~Relative  $\mu_\alpha$& -14.44 $\pm$ 0.13 mas yr$^{-1}$\\
~~~~~~~Relative  $\mu_\delta$&  -190.25 $\pm$ 0.12  mas yr$^{-1}$\\
~~~~~~~$\vec{\mu} = 190.79$ mas  yr$^{-1}$\\
~~~~~~~P.A. = $184\fdg3$\\
{\it HST}: model with reference star EDR3 1 mas and 1 mas$^{-1}$ prior errors\\ 
~~~~~~~Absolute $\varpi$& 64.11 $\pm$ 0.13    mas \\
~~~~~~~Relative  $\mu_\alpha$& -14.38 $\pm$ 0.13 mas yr$^{-1}$\\
~~~~~~~Relative  $\mu_\delta$&  -190.28 $\pm$ 0.12  mas yr$^{-1}$\\
~~~~~~~$\vec{\mu} = 190.83$ mas  yr$^{-1}$\\
~~~~~~~P.A. = $184\fdg3$\\
\G~EDR3 Absolute $\varpi$& 64.09 $\pm$ 0.09     mas \\
~~~~~~~Absolute  $\mu_\alpha$& -15.03 $\pm$ 0.08 mas yr$^{-1}$\\
~~~~~~~Absolute  $\mu_\delta$&  -190.90 $\pm$ 0.07  mas yr$^{-1}$\\
~~~~~~~$\vec{\mu} = 191.49$ mas  yr$^{-1}$\\
~~~~~~~P.A. = $184\fdg5$
\enddata
\end{deluxetable}

\begin{deluxetable}{c r r r}
\tablewidth{0in}
\tablecaption{Component Mass  Upper Limits}   \label{tbl-F-all}
\tablehead{
\colhead{Component}&
 \colhead{$P$ [yr] } &
   \colhead{$\alpha$ [mas]\tablenotemark{a}} &
   \colhead{\m [\mjupe]} 
   		}
\startdata
b&1.8&0.35&4.3\\
d&0.8&0.35&7.0\\
e&10.9&1.20&4.4
\enddata
\tablenotetext{a}{Detectable perturbation size given reference frame noise levels.}

\end{deluxetable}

\begin{deluxetable}{l r r r r r r}
\tablecaption{\HSTs Exoplanet Perturbations and Inclinations\label{tbl-incs}}
\tablewidth{0in}
\tablehead{
\colhead{ID} &  
\colhead{$\alpha$ [mas]} &
\colhead {$i$ [$\arcdeg$]} &
\colhead {$i_{\rm corr}$\tablenotemark{a} [$\arcdeg$]} &
\colhead{$RUWE$\tablenotemark{b}} &
\colhead{$\chi^2$~\tablenotemark{c}} &
\colhead{Source\tablenotemark{d}}
		}
\startdata
$\upsilon$ And d&1.39$\pm$ 0.07&23.8 $\pm$1.3&23.8&7.25&6.39&1\\
$\upsilon$ And c&0.62 0.08&7.9 1&7.9&-& -&1\\
$\gamma$ Cep Ab&1.1 0.1&169.5 1.1&10.5&3.21&4771&2\\
$\epsilon$ Eri b&1.88 0.2&45 8&30.1&2.72&33.89&3\\
HD 33636 A&5 0.2&14 0.1&14&1.88&55.6&4\\
HD 136118 b&1.45 0.25&163.1 3&16.9&1.43&71.32&5\\
GJ 876 b&0.25 0.06&84 6&84&1.34&3.56&6\\
HD 128311 c&0.46 0.09&56 15&56&1.31&12.64&7\\
HD 38529 c&1.05 0.06&48.3 3.7&48.3&1.05&5.3&8\\
HD202206  c&0.76 0.11&7.7 1.1&7.7&1.03&32.25&9\\
Prox Cen c&0.5 0.1&18 4&18&0.97&0.51&10\\
55 Cnc d&1.9 0.4&53 7&53&0.86&1.81&11\\
\mAs &&&&0.86&2.35&
\enddata
\tablenotetext{a}{$i_{\rm corr}=90\arcdeg-{\rm mod}(i,90\arcdeg)$ for $i>90\arcdeg$.}
\tablenotetext{b}{$RUWE$, reduced unit weight error from \Gs EDR3. Larger RUWE implies  photcenter motion in excess of measured parallax and proper motion.}
\tablenotetext{c}{A larger $\chi^2$ value indicates more significant  acceleration in proper motion \citep{Bra21}, thus a higher probability of a perturbing companion.}
\tablenotetext{d}{1=\cite{McA10}, 2=\cite{Ben18}, 3=\cite{Ben22}, 4=\cite{Bea07}, 5=\cite{Mar10}, 6=\cite{Ben02b}, 7=\cite{McA14}, 8=\cite{Ben10}, 9=\cite{Ben17b}, 10=\cite{Ben20b}, 11=\cite{McA04} }

\end{deluxetable}

\begin{deluxetable}{l  r r r r}
\tablecaption{KS Test Results\label{tbl-KSres}}
\tablewidth{5in}
\tablehead{
\colhead{Test} &  
\colhead{D} &
\colhead {C} &
\colhead {$\alpha$} &
\colhead {p} 
		}
\startdata
\HSTs MLR vs 6th Catalog&0.35&0.34&0.050&0.02\\
ExoP vs 6th Catalog&0.53&0.41&0.050&0.00
\enddata
\end{deluxetable}

\clearpage
%
%

\clearpage


\clearpage

\end{document}